# All molecules are entangled in the number of electrons following a general physical principle

Jing Kong*

Department of Chemistry and Center for Computational and Data Sciences, Middle Tennessee State University, 1301 Main St., Murfreesboro, TN 37130, USA

* Correspondence: jing.kong@mtsu.edu

**Abstract**:

A physically motivated equation that determines the number of electrons of a molecule is proposed based on some generally accepted basic chemical knowledge. It permits a complete local definition of molecules and shows that all molecules are entangled in the number of electrons. It also results in the extensivity and convexity of molecular energy, which underpin molecular quantum mechanics. The proposed physical principle includes the molecular size consistency principle as a special case and reveals an intrinsic relationship between molecular locality, size consistency, and energy convexity. Application of wavefunction theory to the new principle shows that an individual molecule with a noninteger number of electrons is locally physical. The energy of a molecule is piecewise linear with respect to its number of electrons. The realness of the number of electrons allows the definition of the electronic chemical potential of a single molecule. A state function equivalent to the energy of a molecule can be defined using the chemical potential as a variable and used to express the physical principle as a simple additivity. The latter also shows that the quantum entanglement in the number of electrons can be viewed as all molecules sharing one chemical potential. When the energy of a molecule with a definite number of electrons is not convex with respect to the number of electrons, the state of the molecule is not always locally determinable even with a known number of electrons.

## Introduction

The predominant basic assumption of chemistry about matter is a classic one, supplemented by quantum mechanics: physical objects are divisible; they are divided into molecules; and individual molecules should be treated with quantum mechanics. On the other hand, quantum entanglement experiments validate another worldview that objects separated at a macroscopic scale can be part of a single wavefunction. This means that all seemingly independent objects must be part of a single wavefunction, because quantum mechanics supersedes classical mechanics. Indeed, the precise one-to-one mapping between electron density and external potential, shown by the Hohenberg-Kohn first theorem [1], implies that no individual molecule can be independent of the other[2]. Those two worldviews are in an apparent conflict, and the contemporary all-inclusive quantum view is still not widely shared due to the effectiveness of the classic view, despite the quantum view being logically sound. The difference between the two views is not merely philosophical, as they, especially the classic one, are part of the basic assumptions underlying molecular research.

Here, we argue that all classically defined molecules are quantum-mechanically entangled in the number of electrons. The argument is in two steps. First, a simple and nontrivial physical principle for molecules can be reasonably derived from a set of observations and beliefs commonly held by molecular scientists in humans' sense of space and time without explicitly invoking quantum mechanics. This set of knowledge, to be listed later in this paper, is referred to as basic chemical knowledge henceforth. It shows that the number of electrons of an individual molecule in reality cannot be determined locally. Instead, it is determined collectively with all the molecules in the universe, i.e., all molecules are entangled in the number of electrons. The equation also leads to the molecular energy convexity that underpins the molecular quantum mechanics. Second, application of wavefunction theory to the principle shows that an individual molecule with a noninteger number of electrons is locally physical even though its number of electrons fluctuates. The energy of a molecule is piecewise linear with respect to its number of electrons. The nonnegative realness of the number of electrons allows the definition of an electronic chemical potential of a single molecule. A state function equivalent to the energy of a molecule can be defined using the chemical potential as a variable. It can be used to express the new physical principle as a simple additivity that shows the extensivity of matter. It also shows that the quantum entanglement in the number of electrons can be viewed as all molecules sharing the same chemical potential. The latter is an interval in general.

We note that the word 'entangled' used here describes a local object that possesses a property undeterminable by locally measurable forces. The term itself is not necessarily associated with quantum theory. We understand that quantum entanglement in literature is often used for delicately arranged experiments or special theoretical conditions associated with the collapse of wavefunctions due to observations. Here we use the word for its literal meaning.

## Previous work: Size-consistency principle

Part of the classic chemical assumption can be represented by the so-called molecular size consistency principle[3]. It states that the total energy[i] of a system made of two molecules $A$ and $B$ separated by a very large (e.g., visible) distance is additive:

$$E(A..B) = E(A) + E(B) \quad (1)$$

[i] Energies of all molecules are assumed to be for ground states and have the same reference point independent of the number of electrons.

We use '..' to indicate a visibly measurable distance. We call this type of separation asymptotic henceforth.

Eq.(1) represents the following basic chemical knowledge:

1. Matter can be separated into molecules. Specifically, it can be divided into molecules with atoms as special cases. Molecules are locally measurable by definition.
2. Molecules have one definite internal energy because chemical reactions produce energy changes. This energy is, of course, relative to a constant reference value for all molecules.
3. Stable objects visibly separated from each other are independent of each other, with the omission of magnetic and gravitational interactions.

Eq.(1) shows the linear extensivity of matter in energy through additivity. It projects properties at the macroscale (i.e., many molecules) to the properties of molecules at the microscale (i.e., individual molecules). The principle is routinely used in molecular quantum mechanics to measure the correctness of approximate methods. Some well-defined approximate quantum methods yield different results when applied to the LHS and the RHS of Eq.(1), indicating their incorrectness due to a lack of size consistency.

Eq.(1) says that molecules in ground states are separable from each other in the language of quantum mechanics. This claim is an approximation from a quantum-mechanical point of view because the electron density, being above zero everywhere [4], is not exactly separable, no matter how far the two sets of nuclei are from each other. This is to say that one cannot go from the LHS to the RHS exactly. Therefore, it is a physically motivated hypothesis. Besides, Eq.(1) is incomplete because it does not specify the number of electrons for each molecule involved, i.e., one cannot go from the RHS to the LHS in general. For example, it would not be correct if A is $H^+$ and B is $H^-$ , because ($H^+$.. $H^-$ ) is not a ground state, and therefore not suitable for the LHS. This indicates that molecules A and B are not independent of each other, although one might assume the opposite from the additive nature of the equation.

**Present work**

We suggest that the following equation be a more comprehensive representation of the locality of a molecule in the classic view:

$$E((X..Y)_L) = \min_N (E(X_N) + E(Y_{L-N})) \tag{2}$$

Here, $(X..Y)_L$ symbolizes a molecule with two asymptotically separated sets of nuclei $X$ and $Y$ and $L$ number of electrons. $X_N$ is a molecule with a set of nuclei $X$ and $N$ number of electrons. Each set of nuclei includes nuclear charges and their relative positions (geometry). $L$ and $N$ are non-negative integers. The minimizer $N^*$ is required to be unique.

Eq.(2) has a simple physical meaning: Two asymptotically separated molecules exchange electrons virtually until their total energy is minimized, with both the minimum energy and the minimizer being unique. It quantitatively reflects the following basic chemical knowledge and assumptions:

4. Molecules are made of atoms, and atoms are made of negatively charged electrons and positively charged nuclei with integer charges, as concluded from J. J. Thomson's experiment in 1897. Individual electrons and nuclei are indivisible. Electrons are transferable between separated molecules accompanied by energy variation.
5. Electrons do not hold stable local structures on their own in reality because there is no stable accumulation of free electrons at the macroscale that we can detect under normal chemical circumstances. Rather, they are always near nuclei spatially, i.e., bound by nuclear potentials. It is also consistent with the singleness of molecular energy. In contrast, nuclei can have stable local structural arrangements because physical objects have visible variations.

6. The number of electrons of a molecule is definite because fluctuations of charges are not observed for stable matter.
7. We initially focus on those molecules that are small enough such that the Coulombic interaction between any pair of electrons is not negligible. This means that the energy of a molecule is nonlinear with respect to its number of electrons, i.e., its ionization potential is different from its electron affinity. Otherwise, we would observe zero net change of energy for electron transfers between separated molecules of the same kind.
8. The variational principle is applicable to calculating the energy of a stable molecule. Note that the equation does not dictate a specific formula for energy.

Eq.(2) represents the above-listed basic chemical knowledge concisely:

- It includes the properties 1-3 that underpin the molecular size consistency principle (Eq.(1)).
- The expression $E(X_L)$ is justified by the singleness of the molecular energy with a definite number of electrons. This is to say that the energy of a molecule is locally definable with a given number of electrons and a set of nuclei.
- The requirement of integer $N$ shows the indivisibility of electrons.
- Minimization is the application of the variational principle.
- The singleness of the minimizer is due to the singleness of molecular energy and the nonlinearity of molecular energy with respect to the number of electrons.
- There is no need for additional terms on the RHS because electrons are always with nuclei spatially.
- It is time-independent due to the stability of molecules.

Significant results may be obtained from Eq.(2). First, unlike the molecular size consistency principle (Eq.(1)), Eq.(2) justifies the complete local definition of a molecule. Eq.(1) is applicable only when $N = N^*$, i.e., the two molecules on the RHS of Eq.(1) are not locally deterministic as they appear. On the other hand, the additivity that one may expect for the extensivity of molecular energy seems missing from the non-linear nature of the RHS of Eq.(2). We demonstrate, towards the end of this paper, that additivity is recovered through an alternative form.

Eq.(2) also shows that all molecules in the universe are entangled in the number of electrons. The molecule on the LHS represents a lone molecule in the universe[ii]. The molecules on the RHS represent individual molecules in the universe and are said to be open because they can exchange electrons with each other. Eq.(2) says that the determination of the local parameter $N$ is non-local. Furthermore, it is independent of the intermolecular distance beyond the asymptotic separation of individual molecules, i.e., the determination is not due to some tiny force from the other molecule. The nonlocality in the determination of the stable $N$ means that the molecule $X_N$ is not locally deterministic. The phenomenon of this global entanglement is not unusual if one thinks of $X$ and $N$ as variables for a collective state of electrons in analogy to the state variables of classical thermodynamics, where open systems are allowed to interact distantly.

[ii] We consider the universe to be made of molecules only and in its ground states. The number of them can be arbitrarily large but finite.

The nonlocality of far-separated molecules has been discussed before in the context of the exchange-correlation potential of density functional theory and charge transfer, e.g., Refs [5, 6]. In practice, the number of electrons of a molecule can often be correctly guessed based on electron affinity and ionization potential data, which is equivalent to a manual application of Eq.(2). Molecules in nature are typically neutral due mainly to the fact that the smallest ionization potential is larger than the largest electron affinity in the periodic table. But some molecules, such as superalkali, have ionization potentials as low as 3.51eV[7], which is lower than the electron affinity of chlorine.

Eq.(2), with the singleness requirement of the minimizer, implies the equivalence of the following three statements:

A. The energy of a molecule is linearly extensible, i.e., $E(A..A)=2E(A)$.
B. The state of a molecule does not change due to the appearance of its copy in another corner of the universe, i.e., an open molecule can be treated as if alone and is allowed on the LHS of Eq.(2).
C. The ionization potential decreases as the number of electrons increases, i.e., $I(X_L)>A(X_L)$. We call this condition molecular energy convexity (MEC).

Statement A is obvious. Statement B is an alternative to Statement A. To prove C from A, let us consider the system $(X..X)_{2K}$ with an integer $K$, which has an energy $2E(X_K)$ per Statement A. Let the opposite of MEC be true, i.e., $E(X_{K-1})+E(X_{K+1})\le 2E(X_K)$ . Then, $N=K$ would be either a local maximum or the system has the same energy for $N=K-1,K,K+1$, both of which violate the single minimum requirement. The derivation here also shows that intrinsic relation between the MEC and the molecular locality.

The MEC assumption was proposed as a prerequisite for the piecewise linearity of the universal density functional [8]. It was suspected to be particular to the Coulomb repulsion between electrons [4, 9] until very recently, when counterexamples were found for specially designed Coulombic [10] and molecular systems [11]. Still, it is a condition that has been assumed in almost every molecular quantum-mechanical calculation without being explicitly stated, because the operator in the Schrödinger equation requires a definite number of electrons for computing $E$. When the MEC does not hold for $E$, the single minimum requirement for Eq.(2) cannot be always maintained. For instance, $N$ is a local maximum if $2E(X_N)>E(X_{N-1})+E(X_{N+1})$ when the LHS of Eq.(2) is $(X..X)_{2N}$. This example also shows that the linear extensivity of molecular energy no longer holds for $E$ because $E(X_N..X_N)=E(X..X)_{2N}$ is smaller than $2E(X_N)$, assuming the molecule $(X..X)_{2N}$ is separable and Eq.(2) is applicable without the single-minimum requirement. In light of the counterexamples to the MEC, a physically reasoned principle like Eq.(2) is needed as an explicit assumption for applying the Schrödinger equation to molecular systems. It is expected to hold for practical problems since the Schrödinger equation has been successfully applied thus far.

Still, Eq.(2) has difficulty in maintaining the strict singleness of the minimizer, notwithstanding the counterexamples to the MEC. To see this, we note that the equation has a single minimizer when the lone minimizer of the equation satisfies the following conditions because of MEC:

$$I(X_{N^*})>A(Y_{L-N^*}),\quad A(X_{N^*})<I(Y_{L-N^*}) \tag{3}$$

But what if $I(X_{N^*}) = A(Y_{L-N^*})$? In this case, $N^* - 1$ also yields the minimum energy. A simple example is $X = \mathrm{H}, Y = \mathrm{H}^+$. So the single minimizer requirement needs to be modified. The updated one is that the lone minimizer is an interval between two consecutive integers, with a length of either 0 or 1. We note that this change does not affect the deduction of the MEC from Eq.(2) above because a strict local maximum is still prohibited.

The assumption of linear extensivity permits the extrapolation of the energy of an individual molecule from the energy of a bulk, i.e., the energy of $X_N$ is the result of an average over a system composed of $K$ sets of the same nuclei $X$ with total $N \times K$ number of electrons in a very thin gas system. A more general averaging would be for the total number of electrons to be an arbitrary integer, rather than an integer multiple of $K$, where a molecule, on average, has a fractional number of electrons. Indeed, this was considered previously as the physical basis for a molecule to have a fractional charge in a seminal work in the context of density functional theory [8, 12]. This extension of the molecular concept allowed $N$ in $X_N$ to be fractional for representing a system of many molecules of the same set of nuclei. The energy of the so-defined molecule $E(X_N)$ was piecewise linear with respect to $N$, based on Janak's theorem and the MEC.

One may also take $E(X_N)$ as the average energy per $X$ without invoking DFT explicitly. Then, Eq.(2) can be formally extended by letting $N$ be real and requiring the minimizer $N^*$ to be either an integer or a real between two consecutive integers. $N^*$ satisfies the following conditions ($\lceil N \rceil$ (ceiling) is the smallest integer not smaller than $N$. $\lfloor N \rfloor$ (floor) is the largest integer not larger than $N$.):

$$I(X_{\lceil N^* \rceil}) \geq A(Y_{\lfloor L-N^* \rfloor}), \quad A(X_{\lfloor N^* \rfloor}) \leq I(Y_{\lceil L-N^* \rceil}) \tag{4}$$

$N^*$ can be a noninteger when the equalities hold. The uniqueness requirement of the minimizer is updated to be either a single integer or a real between two consecutive integers.

The piecewise linearity has been presented and widely accepted as a condition for the exact density functional [8, 13]. However, this line of reasoning was criticized recently [14] with the arguments that DFT was based on wavefunction solutions for individual molecules, and an individual molecule could only have an integer number of electrons physically. We show that the fractional charge for a molecule is a necessary result when wavefunction theory is applied to Eq.(2), assuming that the equation holds for integer molecular charges. While the derivation is standard, it reveals the necessity of a physical principle like Eq.(2).

We start from a case where only inequality occurs in Eq.(4), i.e., the minimization in Eq.(2) is reached with a single integer minimizer. Applying the Schrödinger equation to the whole system $(X..Y)_L$ does not yield the molecules on the RHS exactly because the electron density is above zero everywhere. But it becomes separable at any finite accuracy by approximating the value of the density below a threshold to be zero. If Eq.(2) is correct, the minimizer $N^*$ must match the charge distribution from the global wavefunction to an arbitrary accuracy. Once $N^*$ becomes available, the energy of each molecule on the RHS can be computed with the wavefunction theory as if it were alone per Statement B. Let $\Psi(N^*, X)$ and $\Psi(L - N^*, Y)$ denote the ground state wavefunctions for lone $X_{N^*}$ and $Y_{L-N^*}$, respectively (Possible degeneracies for lone $X_{N^*}$ and $Y_{L-N^*}$ are ignored for now as they do not affect the discussions here.). The ground wavefunction for $(X..Y)_L$ can be accurately approximated as $\mathscr{A}[\Psi(N^*,X)\Psi(L-N^*,Y)]$, as one can verify from Eq.(1). Let $X_{N^*}..Y_{L-N^*}$ symbolize this state. It is the only ground state for $(X..Y)_{N_X+N_Y}$. The corresponding energy $E(X..Y)_L$ can be computed as $E(X_{N^*})+E(Y_{L-N^*})$, following Eq.(1). The minimization of the RHS of Eq.(2) is the application of the variational principle of the wavefunction theory within the trial wavefunction space of $\mathscr{A}[\Psi(N,X)\Psi(L-N,Y)]$ with varying $N$.

When $I(X_{N^*})=A(Y_{L-N^*})$, the LHS of Eq.(2) has two degenerate states, namely $X_{N^*}..Y_{L-N^*}$ and $X_{N^*-1}..Y_{L-N^*+1}$. A superposition of those two states is also a wavefunction for $(X..Y)_L$, which yields a noninteger number of electrons at the locale $X$ as a convex combination of $N^*$ and $N^*-1$, and similarly at the locale $Y$. The energy at the locale $X$ is the same convex combination of $E(X_{N^*})$ and $E(X_{N^*-1})$. This assignment of energy is not arbitrary because it is the only value that is locally definable, required by Eq.(2). It is piecewise linear with respect to the number of electrons.

Physically speaking, the measurement of a property of an individual molecule is on a mixture of all possible local pure states. This mixed state is entangled with the mixed states at other locales as part of the same global wavefunction. In the case of Eq.(2), a ground-state wavefunction for the LHS spans over the two sets of nuclei. The RHS represents the measurement of the energy at locales $X$ and $Y$, respectively, and the detected mixed state at each locale is an ensemble of states with different integer numbers of electrons at that locale. This fluctuation matches the openness of a molecule dictated by Eq.(2). The mixed states of the two open molecules on the RHS are entangled as part of the superposition states for the molecule on the LHS with a definite number of electrons. Furthermore, a wavefunction for the LHS must yield a distribution of electron density that corresponds to the optimized numbers of electrons at those two locales. When $I(X_{N^*})>A(Y_{L-N^*})$, $N^*$ is a single integer, and the mixed state coincides with the pure state for the molecule at the locale as $\Psi(N^*,X)$ because MEC guarantees that the energy of the pure state is lower than any mixed state. The measured number of electrons does not fluctuate. When $I(X_{N^*})=A(Y_{L-N^*})$, $N^*$ is in an interval, $X_{N^*}..Y_{L-N^*}$ is degenerate with respect to the variation of $N^*$ from $\lfloor N^* \rfloor$ to $\lceil N^* \rceil$. The measured number of electrons fluctuates within that interval, and the measurements at the two locales are fully entangled. The measured energy at the locale $X$ changes linearly such that the total energy remains the same with respect to the variation of $N^*$, which matches the behavior of the averaging over many molecules discussed above.

The above derivation is quite standard in molecular quantum mechanics. What is emphasized here is the logical reliance of such a derivation on the physical intuition offered by Eq.(2). Furthermore, the physical reasonableness of the fractional charge of an individual molecule in quantum mechanics comes from the physical reasonableness of Eq.(2) for molecules with integer charges. The possibility of fractional number of electrons is a consequence for molecules being open for electron exchange as stated by Eq.(2) and supported by wavefunction theory. Since all molecules in the physical world are open, individual molecules of noninteger numbers of electrons are just as physically valid as molecules of integer ones. A molecule with non-integer number of electrons differs from one with integer number of electrons in that its number of electrons fluctuates with repeated measurements. Either way, the number of electrons of a molecule must be determined from the global wavefunction, e.g., the wavefunction for $(X..Y)_L$ determines $N^*$ for $X_{N^*}$ and $Y_{L-N^*}$. In short, all molecules are also quantum-mechanically entangled in the number of electrons. We note that an entangled molecule of a non-integer number of electrons is conceptually different from the average of many molecules, even though local measurements in a thin gas cannot distinguish between the two.

To complete Eq.(2), limits related to $N$ for a molecule $X_N$ need to be specified for the optimization. Two obvious conditions are $E(X_N)=\infty$ for $N<0$ and $E(X_0)=0$. A nontrivial consideration is that the number of electrons each set of asymptotically separated nuclei can hold for a bound state is limited. Furthermore, such a limit varies depending on nuclei. Let the limit be symbolized as $N_{\max}$ with the understanding that it is $X$ specific. The limit on the LHS is obviously the sum of the limits on the RHS. To obtain the limit, a formula for $E(X_N)$ needs to be defined for $N>N_{\max}$, which corresponds to unbound states. Furthermore, $E(X_N)$ for $N>N_{\max}$ must satisfy $E(X_N)\geq E(X_{N_{\max}})$ and being convex (albeit not strictly). This ensures the convexity of $E(X_N)$ for all possible $N$ and thus the single minimum for Eq. (2) when the LHS is a bound state. There is no accuracy requirement for the unphysical, unbound states otherwise. Still, this condition is system-specific and thus aesthetically unpleasant. We see a few paragraphs later that the concept of chemical potential can replace the number of electrons as a state variable that has a universal limit for a bound state.

Eq.(2) can be trivially extended to the situation where a molecule on the RHS is also composed of asymptotically separated individual molecules, i.e., the LHS of Eq.(1) can represent such a molecule. The single minimum requirement of Eq.(2) needs to be extended to the lone minimizer being one interval between two integers. The reasoning is that if two molecules (e.g., H and $H^+$) can exchange an electron without changing the total energy of the whole system, two molecular systems made of many of each of those two molecules (e.g., many H's and $H^+$'s, respectively), asymptotically separated from one another, can exchange more than one electron between the two systems without changing the total energy of the whole. The MEC equivalent to the updated requirement on the minimizer includes the possibility of ionization potential and electron affinity being equal, i.e., $I(X_L)\geq A(X_L)$. In this case, the state of $X_L$ can also be an arithmetic average of the states of $X_{L+1}$ and $X_{L-1}$. For instance, the energy of $(p..p)_3$ ('p' symbolizes a proton, '3' means 3 electrons) and an arithmetic average of the energies of $(p..p)_2$ and $(p..p)_4$ are the same. Furthermore, any of the molecules discussed above can be of noninteger charge, which means that the following equation holds:

$$E((X..Y)_N)=\min_{N_X}(E(X_{N_X})+E(Y_{N-N_X})) \quad , \tag{5}$$

with the lone minimizer being an interval. The possibility of a length of the interval larger than one means that the piecewise linear function of $E(X_N)$ may be straight at an integer $N$ where $I(X_N)=A(X_N)$.

Eq.(5) raises a question: Can a molecule with a non-integer number of electrons be alone physically? The answer is No because an electron is indivisible, although the energy of the molecule can be computed as such. Indeed, the separability of molecules shown in the RHS is considered an asymptotic property[2] with wavefunction theory as discussed above. The absolute openness of a molecule with a non-integer number of electrons means that it must be part of a molecule with an integer number of electrons, i.e., it is non-isolable. On the other hand, a molecule with an integer number of electrons can have a dual role: being alone and being open, i.e., isolable.

The realness of the number of electrons allows the definition of the chemical potential of a molecule, symbolized as $\mu$, as a partial derivative of $E(X_N)$ with respect to $N$. It is in analogy to, but independent of, the concept of chemical potential in thermodynamics. Previously, the definition of the chemical potential of an individual molecule in the literature was based on thermodynamics as the motivator[15], which was criticized recently as being arbitrary[16].

Since $E(X_N)$ is convex and piecewise linear for bound states, $\mu$ needs to be defined as a subgradient in general. To fully define $\mu$, $E(X_N)$, which is defined for $N \geq 0$ thus far, needs to be defined for any real $N$. We set $E(X_N)$ to be $+\infty$ for $N < 0$ to effectively exclude $N < 0$ for physical reasons and still maintain the convexity of $E(X_N)$. The value of $\mu$ for a set of nuclei $X$ consistent with Eq.(2) needs to satisfy the following conditions:

$$\mu \in \partial_N E(X_N) = \begin{cases} -\infty, & N < 0 \\ (-\infty, I(X_1)), & N = 0 \\ -I(X_{\lceil N \rceil}), & (0 < N < \lceil N \rceil) \,\&\, (\lceil N \rceil < N_{\max}) \\ (-I(X_N), -A(X_N)), & (N = \lceil N \rceil) \,\&\, (\lceil N \rceil < N_{\max}) \\ (-I(X_{N_{\max}}), x \geq 0), & N = N_{\max} \\ > 0 & N > N_{\max} \end{cases} \tag{6}$$

Qualitative illustrations of $E(X_N)$ and $\mu$ are shown in Figs.1a and 1b, respectively. The lines are straight for either a non-integer $N$ or an integer $N$ with $I(X_N) = A(X_N)$. Due to the convexity of $E(X_N)$, $\mu$ is a monotonically increasing function of $N$, which admits a one-to-one correspondence between $N$ and $\mu$ for a molecule in a broad sense. Therefore, we may use $\mu$ as a state variable instead of $N$, and symbolize a molecule as $X_\mu$. Furthermore, a state function equivalent (conjugate) to $E(X_N)$ can now be defined:

$$G(X_\mu) \stackrel{def}{=} \min_N \{E(X_N) - \mu N\} \tag{7}$$ [iii]

A known $G(X_\mu)$ may be used to obtain $E(X_N)$:

$$E(X_N) = \max_\mu \{G(X_\mu) + \mu N\} \tag{8}$$

The number of electrons corresponding to a $\mu$ can be obtained as $N \in \partial_\mu G(X_\mu)$. $G(X_\mu)$ is piecewise linear with respect to $\mu$ for bound states because of the piecewise linearity of $E(X_N)$, as illustrated in Fig.1c.

[iii] The function being optimized may resemble the Lagrangian for self-consistent-field(SCF) calculations, but the two equations are obviously different. The '$\mu N$' term in SCF is for keeping the number of electrons constant.

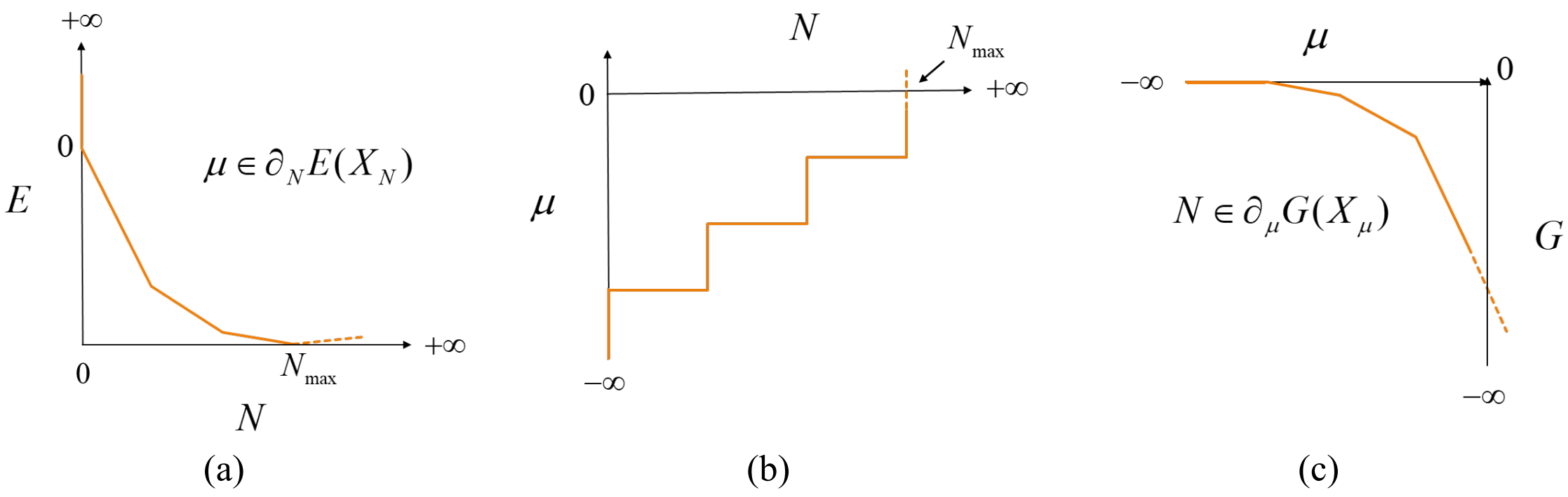

Figure 1. Illustrations of the shapes and relationships among $E(X_N)$, $G(X_\mu)$, $N$ and $\mu$. The dashed segments represent unbound states and are not necessarily linear. (a) $E(X_N)$ versus $N$. (b) $N$ versus $\mu$. (c) $G(X_\mu)$ versus $\mu$.

Eqs.(7) and (8) are defined for $N$ and $\mu$ being all real numbers, respectively, and thus are for both bound and unbound states. However, it is desirable to make the search domains for $E$ and $G$ tighter, especially since Eq.(2) is valid for bound states only. Eq.(6) shows that a finite negative $\mu$ always corresponds to a bound state ( $0 \le N \le N_{\max}$ ) and a bound state is always within the reach of a finite negative $\mu$. Therefore, the searches for bound states can be effectively limited to $N > 0$, $\mu < 0$, and $-\infty < N\mu < 0$, and the conjugation between $E$ and $G$ still holds. The tradeoff is that $N$ cannot be zero, i.e., the universe must hold at least one bounded electron. $G(X_\mu)$ and $E(X_N)$ for bound states have the following formulas, with $N_\mu$ standing for an $N$ corresponding to a given $\mu$ according to Eq.(6):

$$G(X_\mu) = E(X_{\lceil N_\mu \rceil}) - \lceil N_\mu \rceil \mu , \quad E(X_N) = G(X_{-I_{X_{\lceil N \rceil}}}) - I_{X_{\lceil N \rceil}} N \tag{9}$$

One may wonder what can be gained from $\mu$ and $G$. First, a universal condition $\mu < 0$ is sufficient for all bound states. Second, the quantum mechanical uncertainty principle is reflected in the duality between $\mu$ and $N$ per Eq.(6) (See also Fig.1b): When $\mu$ is uncertain (i.e., in an interval), $N$ is an integer (i.e., certain); When $\mu$ is certain, $N$ may be uncertain as a result of an ensemble of states with different numbers of electrons.

Third, one can derive from Eq.(2) the following simple additive rule for $G$ for bound states:

$$G((X..Y)_\mu) = G(X_\mu) + G(Y_\mu) , \quad -\infty < \mu < 0 \ , \tag{10}$$

This equation is obviously extendable to many but finite number of molecules of different atomic structures. Its physical meaning is that all stable molecules in the universe share the same negative chemical potential. This connection between separated molecules reflects quantum entanglement explicitly. The upper and lower bounds of $\mu$ on the LHS are the lowest upper-bound and the highest lower-bound among the molecules on the RHS, respectively. Thus, when a system contains a molecule with one definite chemical potential, the chemical potential of all the molecules equals that of the molecule, i.e., the gap between the bounds is zero. A definite chemical potential is a feature of a fractionally charged molecule, but it can also happen to a molecule with an integer number of electrons, as discussed above. If a system contains two molecules with a definite chemical potential, the one with the larger interval defines the chemical potential for the whole system. On the other hand, if the chemical potential of a system has a non-zero gap, the chemical potential of an individual molecule in the system can vary between the bounds without causing electron transfer. This explains the seeming independence of separated objects at the macroscale. Still, the chemical potential of those molecules is globally determined.

Eq.(10) can also be viewed as a proper principle for molecular size consistency with a simple additivity for the extensivity of matter. This is in contrast to Eq.(1), where the matching of the chemical potential on both sides of the equation is not explicit. Overall, Eq.(10) shows the entanglement of all the molecules in a universe and the extensivity of matter simultaneously.

All the above analyses are not applicable to the case of the theoretical possibility that $E(X_L)$ based on the Schrödinger equation is not convex with respect to $L$. $E(X_L)$ is still physically correct if $X_L$ is alone or isolated. But its physical energy can be lowered in an open environment with freely available electrons as an average over an ensemble of states with various numbers of electrons:

$$\bar{E}(X_L) = \min_{w_i \ge 0, \sum_i w_i = 1, \sum_i w_i L_i = L} (w_i E(X_{L_i})) \tag{11}$$

It has been discussed in the literature as a more general form of molecular energy [17-19]. $\bar{E}$ forms a piecewise-linear convex envelope of $E$ with respect to $L$ by definition, with $\bar{E} = E$ at corners. For an $L$ with $E(X_L)$ above the envelope, $\bar{E}(X_L)$ is a weighted sum of the values of $E$ at the two corners adjacent to $L$ on the envelope (Note that the two corners may differ from $L$ by more than one.). Furthermore, the weights are determined from the positions of those three points. We assume that $E$ is limited to bound states.

Eq.(2) holds for $\bar{E}$ assuming that it holds for $E$ without the constraint of a single minimizer. The assumption is to say that the formulation of $E$ is based on the variational principle and that molecules are assumed to be asymptotically separable from each other. For instance, $E$ can be the solution of Schrödinger equation with $\mathscr{A}[\Psi(N,X)\Psi(L-N,Y)]$ as the basis of the solution space. $\bar{E}((X..Y)_L)$ is then the convex envelope of the RHS of Eq.(2), with possibly multiple minimizers for each $L$ . The same $\bar{E}((X..Y)_L)$ can also be calculated as the minimization of the sum of the convex envelopes of $E(X_N)$ and $E(Y_{L-N})$, showing that Eq.(2) holds for $\bar{E}$. Furthermore, the minimizer is automatically unique, since $\bar{E}(X_N) + \bar{E}(Y_{L-N})$ is convex with respect to both integer $L$ and $N$.

$\bar{E}$ makes obvious physical sense as an average energy per molecule for a bulk system. On the other hand, $E$ is the energy of a lone molecule. The question is whether $\bar{E}$ makes physical sense for an individual molecule in an open environment when $\bar{E} < E$. The answer is that it may, but not always. As an example, let $X$ be an external potential with $E(X_L)$ being strictly convex with respect to integer $L$ except $L = 3$, where $\bar{E}(X_3) < E(X_3)$, i.e., $\bar{E} = E$ for all $L$ except $\bar{E}(X_3) = \frac{1}{2}(E(X_2) + E(X_4))$. Let $Y$ be an external potential with $E(Y_L)$ being convex, with $I(Y_M) = A(Y_M) = 1/2(E(X_2) - E(X_4))$ for a certain $M$. For a lone system $(X..Y)_{M+3}$, the minimizer of the RHS with $\bar{E}$ is the interval [2,4] and the minimized energy is $\bar{E}(X_3) + E(Y_M)$. The superposition of the degenerate wavefunctions corresponding to $(X_2..Y_{M+1})$ and $(X_4..Y_{M-1})$ with equal weights is also a solution of the Schrödinger equation for $(X..Y)_{M+3}$ as a lone molecule, and has 3 electrons near $X$. Therefore, $\overline{E}(X_3)$ makes sense as the energy of a molecule $X_3$. It describes a state of the molecule $X_3$ with a fluctuating number of electrons between 2 and 4. Such a state is non-isolable even though it has an integer number of electrons.

The degeneracy of the lone molecule $(X..Y)_{M+3}$ also means that it is possible for the number of electrons to become fractional near $X$ and $Y$ on the RHS of Eq.(2) with $\overline{E}$, i.e., $N$ may be any value in the interval [2,4] for the molecule $X_N$. This analysis can be trivially generalized to any valid number of electrons. It is the same as the one used above for justifying the possible fractional charge of an individual molecule based on the solution of Schrödinger equation, with $E$ assumed to satisfy the MEC. Thus, $\overline{E}(X_N)$ is applicable for the molecule $X_N$ with a real $N$. It is piecewise linear with respect to $N$ for being the convex envelope of $E(X_N)$. The concept of chemical potential described above is also applicable to $\overline{E}$, and so is Eq.(10).

As one can see, the availability of free electrons of an individual molecule required by $\overline{E}$ is through the degeneracy of the total system. However, such an availability is not always possible. Consider the same molecule $X_3$ as in the above example, but paired asymptotically with a different external potential $Y$ with $E(Y_L)$ being convex with respect to $L$ and $I(Y_M) > E(X_3) - E(X_4)$, $A(Y_M) < E(X_2) - E(X_3)$ for a certain $M$. The molecule $X_3$ as a part of the lone system $(X..Y)_{M+3}$ can only have $E(X_3)$ as its energy because the states of $X_2$ and $X_4$ are not attainable. Indeed, $X_3$ is effectively isolated in this case and must have the energy of the isolated state. A more elaborate example is a lone system $(X..X)_5$. The minimizer on the RHS is the interval [2,3], and $\overline{E}((X..X)_5)$ is $2\overline{E}(X_{2.5}) = 3/2\, E(X_2) + 1/2\, E(X_4)$. This energy expression is impossible to obtain as a solution for a Schrödinger equation of 5 electrons for the lone $(X..X)_5$ because the state of $(X..X)_5$ is either $(X_1..X_4)$ or $(X_2..X_3)$. In either case, its energy is not $2\overline{E}(X_{2.5})$, which requires the availability of both local $X_2$ and $X_4$ states. On the other hand, $(X..X)_5$ may have the energy of $2\overline{E}(X_{2.5})$ as a non-isolable system. In general, a system is non-isolable if the ensemble describing a molecule in a system includes a state that cannot be physically attained without a change of the total energy of the system. This dependence of a molecule's ground state on distant environments, even with a known number of electrons, shows that a molecule's locality is not fundamentally guaranteed.

**Acknowledgement**

The author is grateful to Dr. Emil Proynov for extensive discussions. He also appreciates comments from Dr. Benjamin Janesko.

**References**

1. W. Kohn and L.J. Sham, *Self-consistent equations including exchange and correlation effects.* Phys. Rev. A, 1965, **140**, 1133.
2. J. Kong, *Density functional theory for fractional charge: Locality, size consistency, and exchange-correlation.* The Journal of Chemical Physics, 2024, **161**, 224111.

3. J.A. Pople, J.S. Binkley and R. Seeger, *Theoretical models incorporating electron correlation.* Int. J. Quant. Chem. Symp., 1976, **10**, 1.
4. E.H. Lieb, *Density functional theory for Coulomb systems.* Int. J. Quantum Chem., 1983, **24**, 243.
5. O.V. Gritsenko, R.v. Leeuwen and E.J. Baerends, *Molecular exchange-correlation Kohn–Sham potential and energy density from ab initio first- and second-order density matrices: Examples for XH (X=Li, B, F).* J. Chem. Phys., 1996, **104**, 8535.
6. P.W. Ayers, *On the electronegativity nonlocality paradox.* Theoretical Chemistry Accounts, 2007, **118**, 371.
7. F.A. Cotton, N.E. Gruhn, J. Gu, P. Huang, D.L. Lichtenberger*, et al.*, *Closed-shell molecules that ionize more readily than cesium.* Science, 2002, **298**, 1971.
8. J.P. Perdew, R.G. Parr, M. Levy and J. J. L. Balduz, *Density-functional theory for fractional particle number: Derivative discontinuities of the energy.* Physical Review Letters, 1982, **49**, 1691.
9. P.W. Ayers, *Energy is not a convex function of particle number for r−k interparticle potentials with k > log34.* The Journal of Chemical Physics, 2024, **160**, 044110.
10. S. Di Marino, M. Lewin and L. Nenna, *Ground State Energy Is Not Always Convex in the Number of Electrons.* The Journal of Physical Chemistry A, 2024, **128**, 10697.
11. M.M. González and P.W. Ayers, *Counterexamples to the Convexity of the Energy for Coulomb Particles.* The Journal of Physical Chemistry Letters, 2026, **17**, 2524.
12. J.P. Perdew, *What do the Kohn-Sham orbital energies mean? How do atoms dissociate?*, in *Density functional methods in physics*, J.d. Providencia and R.M. Dreizler, Editors. 1985, Plenum: New York. pp. 265-308.
13. A.J. Cohen, P. Mori-Sanchez and W. Yang, *Insights into current limitations of Density Functional Theory.* Science, 2008, **321**, 792.
14. E.J. Baerends, *On derivatives of the energy with respect to total electron number and orbital occupation numbers. A critique of Janak's theorem.* Molecular Physics, 2020, **118**, e1612955.
15. R.G. Parr and W.Y. Yang, *Density-functional theory of atoms and molecules*. 1989, New York: Oxford University Press.
16. E.J. Baerends, *Chemical potential, derivative discontinuity, fractional electrons, jump of the Kohn–Sham potential, atoms as thermodynamic open systems, and other (mis)conceptions of the density functional theory of electrons in molecules.* Phys. Chem. Chem. Phys., 2022, **24**, 12745.
17. H. Eschrig, *The Fundamentals of Density Functional Theory*. Teubner-Texte zur Physik, Vol. 43. 2003, Stuttgart/Leipzig: Teubner.
18. P.W. Ayers and M. Levy, *Constrained Search in Fock Space: An Alternative Approach to Noninteger Electron Number.* Acta Physico-Chimica Sinica, 2018, **34**, 625.
19. M. Lewin, E.H. Lieb and R. Seiringer, *Universal Functionals in Density Functional Theory*, in *Density Functional Theory: Modeling, Mathematical Analysis, Computational Methods, and Applications*, E. Cancès and G. Friesecke, Editors. 2023, Springer International Publishing: Cham. pp. 115-182.